\documentclass[a4paper,10pt]{article}
\usepackage[latin1]{inputenc}
\usepackage{graphics}
\usepackage{url}
\usepackage{cite}
\usepackage{pdfpages}
\usepackage{longtable}

\begin{document}
\title{Knowledge Ratings in MetaboLights}
\author{
Camila Ramos \and Marco Louro \and Miguel Santos \and Francisco M. Couto
\\
\mbox{ }
\\ \small
LaSIGE, Faculdade de Ci\^encias, Universidade de Lisboa, Portugal
}
%
%
\maketitle
\begin{abstract}
This technical report presents an evaluation of the ontology annotations in the metadata of a subset of entries of MetaboLights, 
a database for Metabolomics experiments and derived information. 

The work includes a manual analysis of the entries and a comprehensive qualitative evaluation of their annotations, 
together with the evaluation guide and its rationale, that was defined and followed.

The approach was also implemented as a software script that given a MetaboLights entry 
returns a quantitative evaluation of the quality of its annotations (available on request).
\end{abstract}
\section{Introduction}
\label{intro}
\subsection{Motivation}
Metabolomics is the systematic study of metabolites in a biological system, whose identification and quantification can provide insights to metabolic processes under certain conditions. These metabolite profiles can potentially act as biomarkers for certain diseases and their applications extend to toxicology, pharmaceutical research and nutrition. Repositories such as MetaboLights play a key role in data sharing and linking resources of that which is by definition a highly integrative field of study~\cite{haug2012metabolights}. Therefore, since it is essential to ensure the quality of annotated information~\cite{couto2014rating}, we selected MetaboLights as our study subject that is exactly our topic of discussion.

\subsection{Problem}
Linking web resources implies using a standardized vocabulary in order to filter out ambiguity and subjectivity. However, it is frequent that this premise is overlooked when articles are submitted, as free text or no information at all is provided instead of ontology terms. It is unlikely that the data in MetaboLights proves to be an exception in this respect, and since there is not a comprehensive evaluation of annotated information in global terms, it is impossible to guarantee its quality beforehand.

\subsection{Objectives}
\begin{enumerate}
\item Produce a global evaluation of the annotations of a set comprising all public studies repositioned in MetaboLights by combining manual and an electronic approaches;
\item Identify and list any specific error or insufficiency found in the metadata file for each analyzed study;
\item Take into account and quantify all instantiated strings of text fielded in the descriptions section in place of ontology terms;
\item Run a statistical analysis of our results in order to support our conclusions regarding the global status of the database;
\item Make suggestions on how to improve the overall quality of stored information, based on our findings.
\end{enumerate}

\section{Framework}
We now present a theoretical framework on information sources, methodologies and technologies we used for this project.

\subsection{Information sources}
Information on the structure and organization of MetaboLights as well as its content, was retrieved from the web page and the corresponding article in May 2015. 
Ontology structure and terms were retrieved directly from BioPortal and EMBL-EBI. 

\subsection{Methodologies and technologies}
Our approach was centered around scoring the entries in MetaboLights. This required the retrieval of all annotations (identified by a PURL) listed in the metadata file for each of the public access studies in MetaboLights. For that, we designed two Python modules; the first one writes the links of the metadata files on a text file and the second one returns a list of PURLs by annotation type and writes all relevant information on a spreadsheet file. We then proceeded to map the appropriate ontology terms using Prot\'eg\'e visualizing tools and web resources.

Each annotation was scored by dividing the terms' depth by the total length of the associated branch (values ranging from 0 to 1). Hence, the more specific a term is, the closer its score is to 1. The type scores were determined by summing up each annotation score and dividing the result by the number of annotations. The global score for an entry was determined by calculating the arithmetic mean of the scores for each type. For this we ruled out all types which had absolutely no annotations in the retrieved files and also study person type (5 studies with annotations). The individual scoring of each annotation was performed manually; for the type scores and global scores a Python module was designed and used. In this process we also took note of any irregularity discovered in the metadata. Additionally, we calculated the log global scores for a more realistic assessment of the quality, since we considered we should favor entries with relatively fewer annotations.

The scores' module also retrieves all unannotated strings under each type. Using the same method, we calculated new type scores taking the number of strings for that type into consideration instead of the number of annotations, which consequently led to different global scores, and compared the results.

\section{Results}
We now present a detailed description of our results as well as examples in which we demonstrate the algorithms we used.

\subsection{Global Evaluation}
The immediate intake from the global scores distribution is the fact that nearly a third (31/95) of the submitted articles have no annotated term (Figure 1). Solely considering the goal of linking information, this represents a substantial flaw in the data. From our perspective, there are two possible solutions for this issue: either encouraging submitters to properly annotate their studies or implement an electronic annotation algorithm (which would be viable since none of the published studies lacks terms fielded as free text and most of them are fairly comprehensible in comparison to their respective ontology terms).

As to information which is in fact annotated, our results show that only a limited number of studies score above a reasonable threshold (log score>70). The comparison between the average scores when considering either the total number of terms (annotated and un-annotated) or the total number of annotations (Table 1) suggests that while the number of free text descriptions has a significant impact on information quality (as it potentially means more of the terms could have been annotated), the existing annotations could possibly be more specific. However, it should also be noted that there are no extremely poor annotations in general terms (log score<30)Be that as it may, this problem can be more easily addressed than the former, using the same methods we suggested and it would improve the quality of annotated information substantially.

Interestingly, the percentage of studies scoring above the mean is higher when we weight by the total number of terms (Table 1). While a lower average value was expected, this result suggests that the issue of unannotated terms is slightly more localized. The same can be inferred from the lower standard deviation value. Nevertheless, from our point of view, it remains a hindrance and should be fixed.

\begin{figure}
\centering
\includegraphics[width=0.9\textwidth]{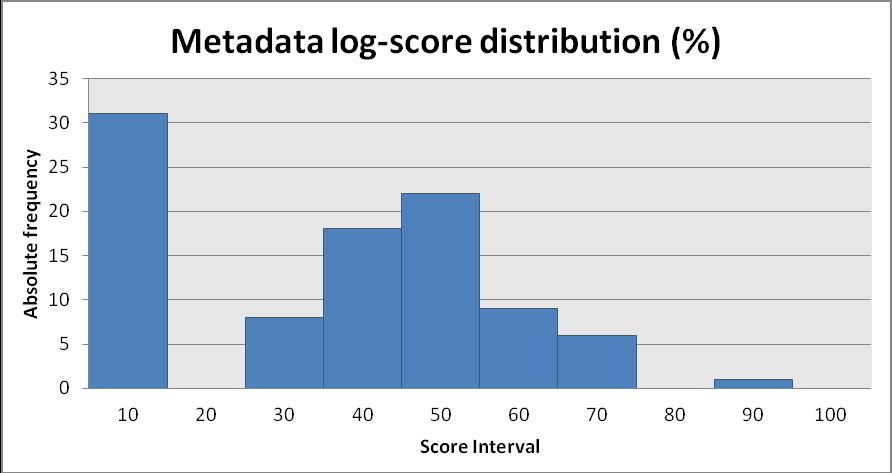}
\caption{Histogram representing the frequency of metadata in each 10 points range. All studies ranking on the [0;10] interval have a score of 0 (no annotations).}
\label{fig1}
\end{figure}

Our observations also identified a general trend in the data in which no submitted study is annotated for all types. Given the overall quality of the annotations, this is not entirely unsurprising. However, it is still noteworthy as even the best scoring studies fail in this aspect and it may reveal some problem the submitters may have in finding the appropriate ontology term.

Another situation which should be addressed is the fact that specific terms and links appear repeated in some metadata files and in consistent fashion. It is unlikely that this is intentional and could possibly be due a bug in the code of the database (as this is only valid for "metabolite profiling" and "mass spectrometry assay") or be a consequence of father-son relationships in the ontologies used.

\begin{table}[!h]
\centering
\resizebox{1.00\textwidth}{!}{ \begin{tabular}{c|cc}
No. of studies: 95 & LogScore(terms) & LogScore(annotations)\\
\hline \hline
Mean & 29,31230073 & 35,18072123 \\
Standard Deviation & 22,62573092 & 27,43255409 \\
Maximum & 80,73549221 & 80,73549221 \\
Minimum(annotated) & 28,54 & 28,54 \\
\% above Mean & 58,94736842 & 53,68421053 \\
\end{tabular} }
\caption{Global statistics for the obtained metadata from the MetaboLights public database. Minimum corresponds to lowest scoring study for which there was at least one annotation.}
\label{tab1}
\end{table}

Regarding our methodology, our choice of disregarding non-PURL links could have had a negative impact on our evaluation of annotation quality, as it would yield undoubtedly higher scores. While we lack statistical support to back this choice, while curating we deemed they were relatively irrelevant. On the other hand, had we not omitted the types for which there was no ontology term in any of the metadata files, scores would drop significantly lower.

\subsection{Type score analysis}
With respect to types scores, assay ranks consistently higher than the others (Figure 2). However, this may be due to the fact that there are only three different annotations in all the files, which is remarkable giving the variety of terms we encountered while searching through several ontologies.

\begin{figure}
\centering
\includegraphics[width=0.9\textwidth]{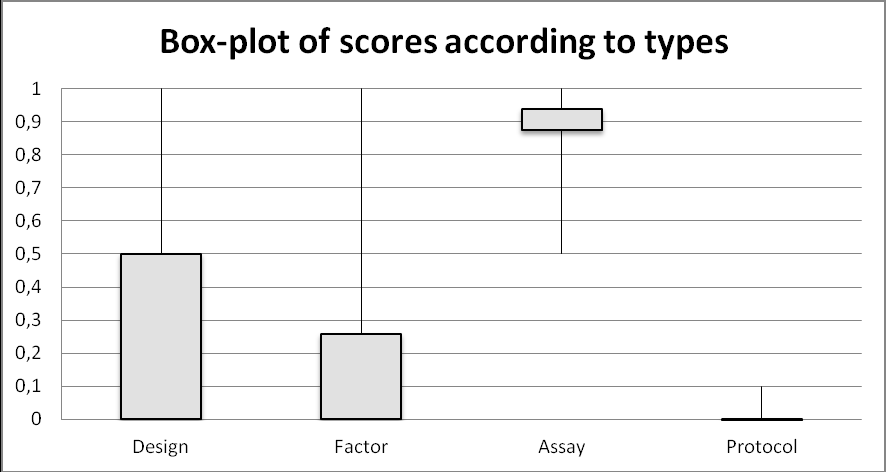}
\caption{Box plot representing the score distribution for each annotated type.}
\label{fig2}
\end{figure}

Protocol on the other hand scores extremely low, which in part can explained by a small number of annotations. We also noted a considerable amount of free-text descriptions containing multiple terms, which can be informative for a curator but can make electronic annotation that more difficult.

Looking at the average score differences (Figure 3) and the score distribution (Figure 2), we can conclude that while design terms are fairly well annotated (in comparison), that type also includes the majority of unannotated terms. As opposed to that, there is no unannotated term under assay. As stated previously, this can be due to the overall redundancy of the annotations.

\begin{figure}
\centering
\includegraphics[width=0.9\textwidth]{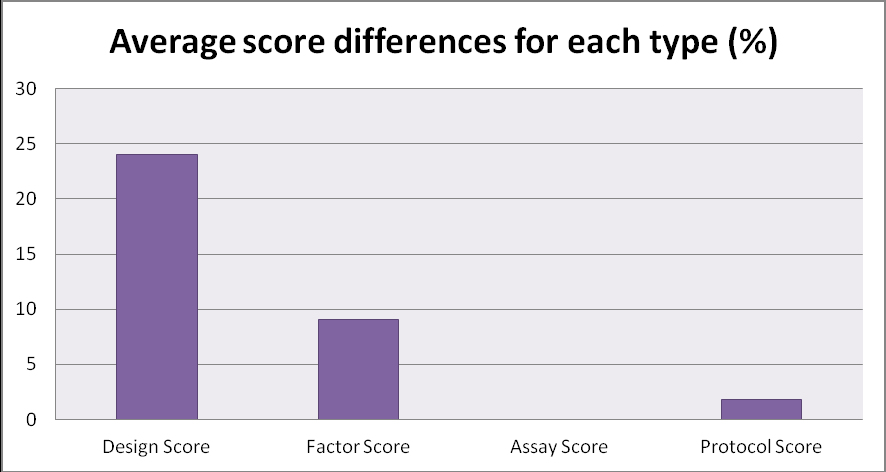}
\caption{Average differences between scores (in percentages) when weighted by the number of annotations or the number of terms requiring annotation. Zero annotations metadata was rejected. Larger differences (16.17\%) are observed in the design score, where there are more terms without any annotation.}
\label{fig3}
\end{figure}

\subsection{Entry-specific considerations}
Over the course of our manual approach, we noticed that entries MTBLS107 through MTBLS111 represented exactly the same study about phytohormones, with each entry pertaining to a specific compound (zeaxanthin, lutein, $\alpha$-carotene, $\beta$-carotene and lycopene, respectively) but otherwise identical in terms of metadata. Given the fact that the annotation regarding the phythormones was unspecific, the intention behind the five submissions might have been a solution to specifying each one. For this and since it is fundamentally the same study, we assert that it is a poor organization of the data, both semantically and concerning the repository itself, and it skews database statistics by attributing extra submissions to the authors' study count.

Additionally, while scoring, we noted the PURL corresponding to the term "ovarian cancer" in MTBLS150 and MTBLS152 was broken ({\tt <error>Ontology not specified or not supported</error}). Our conclusion is that either the authors annotated it incorrectly or the PURL was altered afterwards.

One other situation is that on a few cases terms are listed under two types, but only one of them has the corresponding annotation. This is true for MTBLS81 and MTBLS147; for the former, the term "lipid droplets" is listed under factor type (twice) and design type, with a PURL being provided only for the first one; for the latter, the term "NMR spectroscopy" is listed under assay type and protocol type, and only the term under assay type is annotated. This incongruence is probably due to an error in assigning the string terms, as presumably they should be fielded just under one type, rather than them being unannotated. For example, for all other studies, "NMR spectroscopy" appears under assay type and not protocol type. Also, following the link in MTBLS81, no ontology term was returned ("The page you are looking for wasn't found. Please try again.")

\subsection{Examples}
This section shows the retrieved data for a specific type (identifier: MTBLS95).

Study Design Type:
\begin{itemize}
\item "gas chromatography-mass spectrometry"
\item "Pseudomonas syringae pv. tomato str. DC3000"
\item "Arabidopsis"
\item "type III protein secretion system complex"
\item "MAPK phosphatase export from nucleus"
\item "Metabolomics"
\item "avrPto protein, Pseudomonas syringae"
\end{itemize}

Study Design Type Term Accession Number:
\begin{itemize}
\item \url{http://purl.obolibrary.org/obo/CHMO_0000497}
\item \url{http://purl.obolibrary.org/obo/NCBITaxon_223283}
\item \url{http://purl.obolibrary.org/obo/NCBITaxon_3701}
\item \url{http://purl.obolibrary.org/obo/GO_0030257}
\item \url{http://purl.obolibrary.org/obo/GO_0045208}
\item \url{http://purl.bioontology.org/ontology/MSH/C081695}
\end{itemize}

\begin{table}[!h]
\centering
\resizebox{1.00\textwidth}{!}{ \begin{tabular}{l|ccccc}
& \#Annotations Quality & \#Terms Score (annotations) & Score (terms) \\
\hline \hline
Design & 6 & 5.53 & 7 & 0.9216667 & 0.79 \\
Factor & 0 & 0 & 2 & 0 & 0 \\
Assay & 2 & 1.75 & 2 & 0.875 & 0.875 \\
Protocol & 0 & 0 & 6 & 0 & 0 \\
Score(terms) & \multicolumn{5}{c}{41.625} \\
LogScore(terms) & \multicolumn{5}{c}{50,2075956} \\
Score(annotations) & \multicolumn{5}{c}{44,9166667} \\
LogScore(annotations) & \multicolumn{5}{c}{53,52235268} \\
\end{tabular} }
\caption{Example of the intermediate steps required to calculate the final scores of entry MTBLS95}
\label{tab2}
\end{table}

\section{Discussion}
As per our initial objectives, we feel the result is satisfying. Nonetheless, given the volume of data we gathered, we may need to reorganize some topics covered by this report in order to thoroughly explore and describe our findings, otherwise, as it is now, we would exceed the page limit. Also, some of our groundings lack statistical support, even though we do not perceive this to have an effect on our conclusions.

Perhaps we should have made to our scoring method, by allowing scores to be enriched by sheer number of terms, since the best scoring study only contains four annotations. However, we found no correlation between the number of terms and global scores and that should be explored first. Another aspect which should be analyzed, given the differences between type scores, is the possibility to assign different weights to each one.

Maybe the most interesting prospect we have is to perform a qualitative evaluation of the free-text descriptions and, eventually, an algorithm for electronic annotation if that does not prove to be overly ambitious.

One final suggestion for the database developers: a scoring method similar to that which we used could be employed to block submissions rating under a predetermined value. We feel this would encourage submitters to annotate their studies properly.

\setlength\LTleft{-0.5in}
\setlength\LTright{-0.5in plus 1 fill}
\footnotesize
\begin{longtable}{c|ccccc}
Study Identifier & Total annotations & Score (S) & Log score (S) & Score (annotations) & Log Score (annotations) \\
\hline \hline
MTBLS114 & 4 & 75 & 80,73549221 & 75 & 80,73549221 \\
 MTBLS113 & 12 & 61,33333333 & 69,00445468 & 64,525 & 71,83068219 \\
 MTBLS87 & 7 & 57,9625 & 65,9582106 & 61,2375 & 68,91873194 \\
 MTBLS20 & 6 & 54,95833333 & 63,18803421 & 66,54166667 & 73,58831669 \\
 MTBLS112 & 5 & 54,95833333 & 63,18803421 & 67,45833333 & 74,38021716 \\
 MTBLS166 & 5 & 54,16666667 & 62,44908649 & 75 & 80,73549221 \\
 MTBLS88 & 6 & 53,5 & 61,82386556 & 56,875 & 64,96154591 \\
 MTBLS107 & 4 & 46,875 & 55,45888517 & 46,875 & 55,45888517 \\
 MTBLS108 & 4 & 46,875 & 55,45888517 & 46,875 & 55,45888517 \\
 MTBLS109 & 4 & 46,875 & 55,45888517 & 46,875 & 55,45888517 \\
 MTBLS110 & 4 & 46,875 & 55,45888517 & 46,875 & 55,45888517 \\
 MTBLS111 & 4 & 46,875 & 55,45888517 & 46,875 & 55,45888517 \\
 MTBLS123 & 5 & 46,66666667 & 55,2541023 & 46,66666667 & 55,2541023 \\
 MTBLS52 & 6 & 45 & 53,60529002 & 45 & 53,60529002 \\
 MTBLS119 & 5 & 42,79166667 & 51,3911786 & 42,79166667 & 51,3911786 \\
 MTBLS95 & 8 & 41,625 & 50,2075956 & 44,91666667 & 53,52235268 \\
 MTBLS85 & 5 & 40,21875 & 48,76792789 & 45,8125 & 54,41144022 \\
 MTBLS96 & 6 & 39,66666667 & 48,19877432 & 63,04166667 & 70,52407044 \\
 MTBLS71 & 5 & 39,375 & 47,8971805 & 45,20833333 & 53,81242503 \\
 MTBLS77 & 6 & 39,28571429 & 47,80472968 & 50 & 58,49625007 \\
 MTBLS127 & 4 & 38,54166667 & 47,03199348 & 71,875 & 78,13597135 \\
 MTBLS163 & 7 & 38,4375 & 46,92347937 & 41,75 & 50,33487352 \\
 MTBLS128 & 6 & 37,35833333 & 45,79444393 & 44,31875 & 52,9258748 \\
 MTBLS154 & 6 & 37,16666667 & 45,59299299 & 58,04166667 & 66,03049658 \\
 MTBLS81 & 8 & 37 & 45,41758932 & 57,175 & 65,23717631 \\
 MTBLS170 & 4 & 36,45833333 & 44,84605008 & 43,75 & 52,35619561 \\
 MTBLS90 & 5 & 36,17261905 & 44,54366421 & 67,75 & 74,63127664 \\
 MTBLS165 & 4 & 36,04166667 & 44,40485859 & 43,125 & 51,72756932 \\
 MTBLS26 & 4 & 35,9375 & 44,29434958 & 48,4375 & 56,98556083 \\
 MTBLS147 & 4 & 35,5 & 43,82928516 & 52 & 60,40713237 \\
 MTBLS144 & 5 & 34,375 & 42,62647547 & 42,70833333 & 51,30695822 \\
 MTBLS137 & 3 & 34,375 & 42,62647547 & 46,875 & 55,45888517 \\
 MTBLS93 & 5 & 34,08928571 & 42,31939646 & 67,75 & 74,63127664 \\
 MTBLS157 & 5 & 33,525 & 41,71098842 & 41,29166667 & 49,86763785 \\
 MTBLS92 & 5 & 33,25 & 41,4135533 & 63,0625 & 70,5425039 \\
 MTBLS126 & 5 & 32,70833333 & 40,82589666 & 43,54166667 & 52,14695771 \\
 MTBLS55 & 5 & 32,58928571 & 40,69641987 & 46,875 & 55,45888517 \\
 MTBLS175 & 4 & 32,3125 & 40,39493642 & 42,75 & 51,34907456 \\
 MTBLS155 & 5 & 31,58333333 & 39,59767654 & 41,29166667 & 49,86763785 \\
 MTBLS146 & 5 & 30,625 & 38,54310372 & 53,125 & 61,47098441 \\
 MTBLS79 & 4 & 30,375 & 38,26672527 & 46,5 & 55,09006646 \\
 MTBLS125 & 4 & 29,875 & 37,71237491 & 41,875 & 50,46203924 \\
 MTBLS3 & 4 & 29,6875 & 37,50394313 & 35,9375 & 44,29434958 \\
 MTBLS103 & 4 & 29,16666667 & 36,92338097 & 29,16666667 & 36,92338097 \\
 MTBLS75 & 7 & 28,43181818 & 36,10026654 & 32,55 & 40,65366702 \\
 MTBLS178 & 3 & 28,125 & 35,75520046 & 46,875 & 55,45888517 \\
 MTBLS118 & 3 & 28,125 & 35,75520046 & 40,625 & 49,18530963 \\
 MTBLS117 & 3 & 26,75 & 34,19857472 & 41,375 & 49,95270242 \\
 MTBLS74 & 3 & 25,9375 & 33,27079336 & 48,4375 & 56,98556083 \\
 MTBLS143 & 3 & 25 & 32,19280949 & 46,875 & 55,45888517 \\
 MTBLS131 & 2 & 25 & 32,19280949 & 25 & 32,19280949 \\
 MTBLS156 & 2 & 25 & 32,19280949 & 25 & 32,19280949 \\
 MTBLS104 & 2 & 25 & 32,19280949 & 25 & 32,19280949 \\
 MTBLS116 & 6 & 24,08333333 & 31,13093481 & 29,875 & 37,71237491 \\
 MTBLS2 & 2 & 23,4375 & 30,37807482 & 23,4375 & 30,37807482 \\
 MTBLS86 & 2 & 23,4375 & 30,37807482 & 23,4375 & 30,37807482 \\
 MTBLS150 & 3 & 21,875 & 28,54022189 & 21,875 & 28,54022189 \\
 MTBLS152 & 3 & 21,875 & 28,54022189 & 21,875 & 28,54022189 \\
 MTBLS124 & 2 & 21,875 & 28,54022189 & 21,875 & 28,54022189 \\
 MTBLS148 & 2 & 21,875 & 28,54022189 & 21,875 & 28,54022189 \\
 MTBLS30 & 2 & 21,875 & 28,54022189 & 21,875 & 28,54022189 \\
 MTBLS37 & 2 & 21,875 & 28,54022189 & 21,875 & 28,54022189 \\
 MTBLS39 & 2 & 21,875 & 28,54022189 & 21,875 & 28,54022189 \\
 MTBLS45 & 2 & 21,875 & 28,54022189 & 21,875 & 28,54022189 \\
 MTBLS1 & 0 & 0 & 0 & 0 & 0 \\
 MTBLS10 & 0 & 0 & 0 & 0 & 0 \\
 MTBLS17 & 0 & 0 & 0 & 0 & 0 \\
 MTBLS19 & 0 & 0 & 0 & 0 & 0 \\
 MTBLS21 & 0 & 0 & 0 & 0 & 0 \\
 MTBLS23 & 0 & 0 & 0 & 0 & 0 \\
 MTBLS24 & 0 & 0 & 0 & 0 & 0 \\
 MTBLS25 & 0 & 0 & 0 & 0 & 0 \\
 MTBLS28 & 0 & 0 & 0 & 0 & 0 \\
 MTBLS29 & 0 & 0 & 0 & 0 & 0 \\
 MTBLS31 & 0 & 0 & 0 & 0 & 0 \\
 MTBLS32 & 0 & 0 & 0 & 0 & 0 \\
 MTBLS33 & 0 & 0 & 0 & 0 & 0 \\
 MTBLS34 & 0 & 0 & 0 & 0 & 0 \\
 MTBLS35 & 0 & 0 & 0 & 0 & 0 \\
 MTBLS36 & 0 & 0 & 0 & 0 & 0 \\
 MTBLS38 & 0 & 0 & 0 & 0 & 0 \\
 MTBLS4 & 0 & 0 & 0 & 0 & 0 \\
 MTBLS46 & 0 & 0 & 0 & 0 & 0 \\
 MTBLS47 & 0 & 0 & 0 & 0 & 0 \\
 MTBLS56 & 0 & 0 & 0 & 0 & 0 \\
 MTBLS57 & 0 & 0 & 0 & 0 & 0 \\
 MTBLS59 & 0 & 0 & 0 & 0 & 0 \\
 MTBLS6 & 0 & 0 & 0 & 0 & 0 \\
 MTBLS60 & 0 & 0 & 0 & 0 & 0 \\
 MTBLS61 & 0 & 0 & 0 & 0 & 0 \\
 MTBLS67 & 0 & 0 & 0 & 0 & 0 \\
 MTBLS69 & 0 & 0 & 0 & 0 & 0 \\
 MTBLS72 & 0 & 0 & 0 & 0 & 0 \\
 MTBLS8 & 0 & 0 & 0 & 0 & 0 \\
 \caption{Global scores for all public data from MetaboLights}
\label{tab3}
\end{longtable}
 
\section*{Acknowledgements}
We would like to thank the MetaboLights team (namely, Janna Hastings) and the linkedISA team (namely, Alejandra Gonzalez-Beltran) for following the contributions made during this project. This work was supported by FCT through funding of the LaSIGE Research Unit, ref. UID/CEC/00408/2013.

\bibliographystyle{plain}
\bibliography{references}
%
%
%

\end{document}